\documentclass[12pt]{article}
\input{psfig.sty}
\begin{document}
\begin{center}

\vbox{\vspace{10ex}}

{\LARGE \bf  Lorentz Group in Feynman's World}\\[2mm]

\vspace{4ex}

Y. S. Kim\footnote{electronic address: yskim@physics.umd.edu}\\
{\it Department of Physics, University of Maryland, \\
College Park, Maryland 20742, U.S.A.}

\vspace{4ex}

Marilyn E. Noz\footnote{electronic address: noz@nucmed.med.nyu.edu}\\
{\it Department of Radiology, New York University, \\
New York, New York 10016, U.S.A.}

\end{center}

\vspace{4ex}

\begin{abstract}

R. P. Feynman was quite fond of inventing new physics.  It is shown
that some of his physical ideas can be supported by the mathematical
instruments available from the Lorentz group.  As a consequence, it is
possible to construct a Lorentz-covariant picture of the parton model.
It is shown first how the Lorentz group can be used for studying
the internal space-time symmetries of relativistic particles.  These
symmetries are dictated by Wigner's little groups, whose transformations
leave the energy-momentum four-vector of a given particle invariant.
The symmetry of massive particles is like the three-dimensional
rotation group, while the symmetry of massless particles is locally
isomorphic to the two-dimensional Euclidean group.  It is noted that
the $E(2)$-like symmetry of massless particles can be obtained as an
infinite-momentum and/or zero-mass limit of the $O(3)$-like little
group for massive particles.
It is shown that the formalism can be extended to cover relativistic
particles with space-time extensions, such as heavy ions and hadrons
in the quark model.  It is possible to construct representations of
the little group using harmonic oscillators, which Feynman {\it et al.}
used for studying relativistic extended hadrons.  This oscillator
formalism allows us to show that Feynman's parton model is a
Lorentz-boosted quark model.  The formalism also allows us to explain
in detail Feynman's rest of the universe which is contained in his
parton picture.

\end{abstract}

\newpage

\section{Introduction}\label{intro}

The role of Lorentz covariance in quantum field theory and thus
Feynman diagrams is well known.  In this paper, we study how the
Lorentz group plays roles in other physical theories initiated by
Feynman.
Richard Feynman and Eugene Wigner left their own legacies in physics.
Their approaches to physics appear to be different.  Feynman knew how
to observe the real world and wrote down systematically how the physical
world behaves, while Wigner was able to develop mathematical tools
suitable to physics.

The most controversial phenomenological observation
Feynman made was his parton model.  Among the many contributions
Wigner made, his 1939 paper on the inhomogeneous Lorentz group is still
controversial in that its physical contents are still being explored.
In this report, we propose to combine Feynman's parton picture with
Wigner's representation of the Lorentz group.  The paper thus consists
of two parts.  First, we explain in detail Wigner's representation of
the Lorentz group for internal space-time symmetries of relativistic
particles.  We then deal with Feynman's world using coupled harmonic
oscillators.  It is shown that his parton model can be explained within
the framework of Wigner's representation.

In order to see which aspect of Wigner's work is relevant to Feynmnan's
parton model, let us go back to Einstein.
If the momentum of a particle is much smaller than its mass, the
energy-momentum relation is $E = p^{2}/2m + mc^{2}$.  If the momentum
is much larger than the mass, the relation is $E = cp$.  These two
different relations can be combined into one covariant formula
$E = \sqrt{m^{2} + p^{2}}$.  This aspect of Einstein's $E = mc^{2}$
is also well known.

In addition, particles have internal space-time variables.  Massive
particles have spins while massless particles have their helicities
and gauge degrees of freedom.
As a ``further content'' of Einstein's $E = mc^{2}$, we shall discuss
that the internal space-time structures of massive and massless particles
can be unified into one covariant package, as $E = \sqrt{m^{2} + p^{2}}$
does for the energy-momentum relation.

The mathematical framework of
this program was developed by Eugene Wigner in 1939~\cite{wig39}.
He constructed the maximal subgroups of the Lorentz group whose
transformations will leave the four-momentum of a given particle
invariant.  These groups are known as Wigner's little groups.
Thus, the transformations of the little groups change the internal
space-time variables of the particle, while leaving its four-momentum
invariant.  The little group is a covariant entity and takes different
forms for particles moving with different speeds.

In order to achieve the zero-mass and/or infinite-momentum limit of
the $O(3)$-like little group to obtain the $E(2)$-like little group,
we use the group contraction technique introduced by Inonu and
Wigner~\cite{inonu53}, who obtained the $E(2)$ group by taking
a flat-surface approximation of a spherical surface at the
north pole.  In 1987, Kim and Wigner~\cite{kiwi87jm} observed that
it is also possible to make a cylindrical approximation of the
spherical surface around the equatorial belt.
While the correspondence between $O(3)$ and the $O(3)$-like
little group is transparent, the $E(2)$-like little group contains
both the $E(2)$ group and the cylindrical group~\cite{kiwi90jm}.
We study this aspect in detail in this report.

Let us look at the world of R. P. Feynman.  When we read his papers,
he totally avoids group theoretical languages.  However, whenever
appropriate, his physical reasoning is consistent with the Lorentz
group.  Let us look at Feynman diagrams.  They are
surprisingly consistent with Lorentz covariance of the S-matrix
formalism of quantum field theory.  This aspect is well known.

In this paper, we study Feynman's papers published in
1969~\cite{fey69} and 1971~\cite{fkr71}, and the chapter on density
matrix in his 1972 book on statistical mechanics~\cite{fey72}.
Feynman appears to be dealing with three different physical problems
in these three papers.  However, the Lorentz group allows us to
combine them into one great piece of work which includes a covariant
description of Feynman's parton picture.

How do we do this?  In their 1971 paper~\cite{fkr71}, Feynman
{\it et al.} used harmonic oscillators to work out hadronic mass
spectra.  Even though they used relativistic oscillators, they did
not address the question of whether their formalism constitute a
representation of Wigner's little group.
On the other hand, Wigner did not use harmonic oscillators
too often in his papers, and Feynman did not pay enough attention to
Lorentz covariance.  Thus, in order to combine Feynman's initiative
with Wigner's formalism, we can construct representations of the
little group using harmonic oscillators.  In so doing, we construct
harmonic oscillator wave functions which can be Lorentz-boosted.

The question then is whether we can produce new physics
by combining them.  In this paper, we develop Wigner's mathematical
formalism first, and we then use it to interpret Feynman's physics.

In Sec.~\ref{littleg}, we present a brief history of applications of
the little groups to internal space-time symmetries of relativistic
particles.  It is pointed out in Sec.~\ref{gauge} that the
translation-like transformations of the $E(2)$-like little group
corresponds to gauge transformations.

In Sec.~\ref{o3e2}, we discuss the contraction of the
three-dimensional rotation group to the two-dimensional Euclidean
group.  In Sec.~\ref{contrac}, we discuss the little group for a
massless particle as the infinite-momentum and/or zero-mass limit
of the little group for a massive particle.

In Sec.~\ref{feynm}, we move into to the world of R. P. Feynman.
Feynman was particularly fond of harmonic oscillators in formulating
new ideas.  It is pointed out that harmonic oscillators embrace many
useful properties of the Lorentz group.
In Sec.~\ref{coupled}, we review the quantum mechanics of coupled
harmonic oscillators in which one of them corresponds to the world in
which we do physics, and the other is considered as the rest of the
universe.  In Sec.~\ref{restof}, it is shown that the time-separation
variable in a two-body bound state belongs to Feynman's rest of the
universe.  It is shown also that Feynman's oscillator formalism
includes this time-separation variable.
We show in Sec.~\ref{par} that this $O(1,1)$ formalism enables to
construct a covariant model of relativistic extended particles.  As a
consequence, we show that the quark and parton model are two different
aspects of one covariant object.  It is shown also that this parton
picture exhibits the decoherence effect.

From the historical point of view, we are dealing here with further
contents of Einstein's energy-momentum relation.  This question is
addressed in Sec.~\ref{further}.

\section{Wigner's Little Groups}\label{littleg}

The Poincar\'e group is the group of inhomogeneous Lorentz
transformations, namely Lorentz transformations preceded or followed
by space-time translations.  In order to study this group, we have to
understand first the group of Lorentz transformations, the group of
translations, and how these two groups are combined to form the
Poincar\'e group.  The Poincar\'e group is a semi-direct product of
the Lorentz and translation groups.  The two Casimir operators of
this group correspond to the (mass)$^{2}$ and (spin)$^{2}$ of a given
particle.  Indeed, the particle mass and its spin magnitude are
Lorentz-invariant quantities.

The question then is how to construct the representations of the
Lorentz group which are relevant to physics.  For this purpose,
Wigner in 1939 studied the subgroups of the Lorentz group whose
transformations leave the four-momentum of a given free particle
invariant~\cite{wig39}.  The maximal subgroup of the Lorentz group
which leaves the four-momentum invariant is called the little group.
Since the little group leaves the four-momentum invariant, it governs
the internal space-time symmetries of relativistic particles.
Wigner shows in his paper that the internal space-time symmetries of
massive and massless particles are dictated by the $O(3)$-like and
$E(2)$-like little groups respectively.

The $O(3)$-like little group is locally isomorphic to the
three-dimensional rotation group, which is very familiar to us.
For instance, the group $SU(2)$ for the electron spin is an
$O(3)$-like little group.  The group $E(2)$ is the Euclidean group
in a two-dimensional space, consisting of translations and rotations
on a flat surface.  We are performing these transformations everyday
on ourselves when we move from home to school.  The mathematics of
these Euclidean transformations are also simple.  However, the group
of these transformations are not well known to us.  In Sec.~\ref{o3e2},
we give a matrix representation of the $E(2)$ group.

The group of Lorentz transformations consists of three boosts and
three rotations.  The rotations therefore constitute a subgroup of
the Lorentz group.  If a massive particle is at rest, its four-momentum
is invariant under rotations.  Thus the little group for a massive
particle at rest is the three-dimensional rotation group.  Then what is
affected by the rotation?  The answer to this question is very simple.
The particle in general has its spin.  The spin orientation is going
to be affected by the rotation!

If the rest-particle is boosted along the $z$ direction, it will pick
up a non-zero momentum component.  The generators of the $O(3)$ group
will then be boosted.  The boost will take the form of conjugation by
the boost operator.  This boost will not change the Lie algebra of the
rotation group, and the boosted little group will still leave the
boosted four-momentum invariant.  We call this the $O(3)$-like little
group.

We realize that the standard four-vector convention is $(t, x, y, z)$,
but it is more convenient to use $(x, y, z, t)$ when we study
light-cone coordinate system and group contractions.  In this
non-standard convention, the four-momentum vector for the particle
at rest is $(0, 0, 0, m)$, and the three-dimensional rotation group
leaves this four-momentum invariant.  This little group is generated
by
\begin{equation}\label{j3}
J_{1} = \pmatrix{0&0&0&0\cr0&0&-i&0\cr0&i&0&0\cr0&0&0&0} , \qquad
J_{2} = \pmatrix{0&0&i&0\cr0&0&0&0\cr-i&0&0&0\cr0&0&0&0} , \qquad
J_{3} = \pmatrix{0 & -i & 0 & 0 \cr i & 0 & 0 & 0
\cr 0 & 0 & 0 & 0 \cr 0 & 0 & 0 & 0} ,
\end{equation}
which satisfy the commutation relations:
\begin{equation}
 [J_{i}, J_{j}] = i\epsilon_{ijk} J_{k} .
\end{equation}

It is not possible to bring a massless particle to its rest frame.
In his 1939 paper~\cite{wig39}, Wigner observed that the little group
for a massless particle moving along the $z$ axis is generated by the
rotation generator around the $z$ axis, namely $J_{3}$ of Eq.(\ref{j3}),
and two other generators which take the form
\begin{equation}\label{n1n2}
N_{1} = \pmatrix{0 & 0 & -i & i \cr 0 & 0 & 0 & 0
\cr i & 0 & 0 & 0 \cr i & 0 & 0 & 0} ,  \qquad
N_{2} = \pmatrix{0 & 0 & 0 & 0 \cr 0 & 0 & -i & i
\cr 0 & i & 0 & 0 \cr 0 & i & 0 & 0} .
\end{equation}
If we use $K_{i}$ for the boost generator along the i-th axis, these
matrices can be written as
\begin{equation}
N_{1} = K_{1} - J_{2} , \qquad N_{2} = K_{2} + J_{1} ,
\end{equation}
with
\begin{equation}
K_{1} = \pmatrix{0&0&0&i\cr0&0&0&0\cr0&0&0&0\cr i&0&0&0} ,  \qquad
K_{2} = \pmatrix{0&0&0&0\cr0&0&0&i\cr0&0&0&0\cr0&i&0&0} .
\end{equation}
The generators $J_{3}, N_{1}$ and $N_{2}$ satisfy the following set
of commutation relations.
\begin{equation}\label{e2lcom}
[N_{1}, N_{2}] = 0 , \qquad [J_{3}, N_{1}] = iN_{2} , \qquad
[J_{3}, N_{2}] = -iN_{1} .
\end{equation}
In Sec.~\ref{o3e2}, we discuss the generators of the $E(2)$ group.
They are $J_{3}$ which generates rotations around the $z$ axis, and
$P_{1}$ and $P_{2}$ which generate translations along the $x$ and $y$
directions respectively.  If we replace $N_{1}$ and $N_{2}$ by $P_{1}$
and $P_{2}$, the above set of commutation relations becomes the set
given for the $E(2)$ group given in Eq.(\ref{e2com}).  This is the
reason why we say the little group for massless particles is
$E(2)$-like.  Very clearly, the matrices $N_{1}$ and $N_{2}$ generate
Lorentz transformations.

It is not difficult to associate the rotation generator $J_{3}$ with
the helicity degree of freedom of the massless particle.   Then what
physical variable is associated with the $N_{1}$ and $N_{2}$ generators?
Indeed, Wigner was the one who discovered the existence of these
generators, but did not give any physical interpretation to these
translation-like generators.  For this reason, for many years, only
those representations with the zero-eigenvalues of the $N$ operators
were thought to be physically meaningful representations~\cite{wein64}.
It was not until 1971 when Janner and Janssen reported that the
transformations generated by these operators are gauge
transformations~\cite{janner71,kuper76,kim97poz}.  The role of this
translation-like transformation has also been studied for spin-1/2
particles, and it was concluded that the polarization of neutrinos
is due to gauge invariance~\cite{hks82,kim97min}.

Another important development along this line of research is the
application of group contractions to the unifications of the two
different little groups for massive and massless particles.
We always associate the three-dimensional rotation group with a spherical
surface.  Let us consider a circular area of radius 1 kilometer centered
on the north pole of the earth.  Since the radius of the earth is more
than 6,450 times longer, the circular region appears flat.  Thus, within
this region, we use the $E(2)$ symmetry group for this region.  The
validity of this approximation depends on the ratio of the two radii.

\begin{figure}[thb]  
\centerline{\psfig{figure=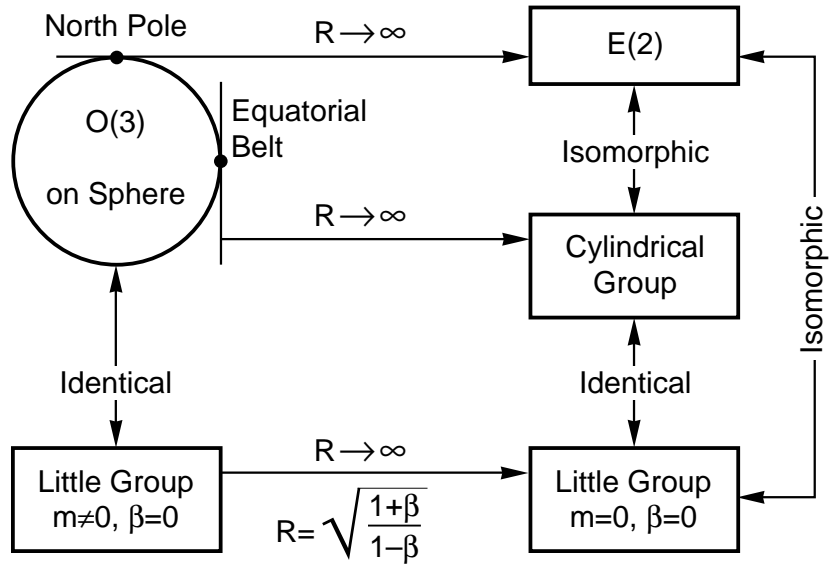,angle=0,height=80mm}}
\caption{Contraction of O(3) to E(2) and to the cylindrical group,
and contraction of the O(3)-like little group to the E(2)-like
little group.  The correspondence between E(2) and the E(2)-like
little group is isomorphic but not identical.  The cylindrical group
is identical to the E(2)-like little group.  The Lorentz boost of
the O(3)-like little group for a massive particle is the same as
the contraction of O(3) to the cylindrical group.} \label{isomor}
\end{figure}

In 1953, Inonu and Wigner formulated this problem as the contraction of
$O(3)$ to $E(2)$~\cite{inonu53}.  How about then the little groups which
are isomorphic to $O(3)$ and $E(2)$?  It is reasonable to expect that the
$E(2)$-like little group be obtained as a limiting case for of the
$O(3)$-like little group for massless particles.  In 1981, it was
observed by Ferrara and Savoy that this limiting process is the Lorentz
boost \cite{ferrara81}.  In 1983, using the
same limiting process as that of Ferrara and Savoy, Han {\it et al}
showed that transverse rotation generators become the generators of
gauge transformations in the limit of infinite momentum and/or zero mass
\cite{hks83pl}.  In 1987, Kim and Wigner showed that the little group for
massless particles is the cylindrical group which is isomorphic to the
$E(2)$ group~\cite{kiwi87jm}.  This is illustrated in Fig.~\ref{isomor}.

\section{Translations and Gauge Transformations}\label{gauge}
It is possible to get the hint
that the $N$ operators generate gauge transformations from Weinberg's
1964 papers~\cite{wein64,hks82}.  But it was not until 1971 when
Janner and Janssen explicitly demonstrated that they generate gauge
transformations~\cite{janner71,kuper76}.  In order to fully appreciate
their work, let us compute the transformation matrix
\begin{equation}
 \exp{(-i(uN_{1} + vN_{2})}
\end{equation}
generated by $N_{1}$ and $N_{2}$.  Then the four-by-four matrix takes
the form
\begin{equation}\label{trans}
\pmatrix{1 & 0 &-u & u \cr 0 & 1 & -v & v \cr
u & v & 1 - (u^{2} + v^{2})/2 & (u^{2} + v^{2})/2 \cr
u & 0 & -(u^{2} + v^{2})/2 & 1 + (u^{2} + v^{2})/2} .
\end{equation}
If we apply this matrix to the four-vector to the four-momentum vector
\begin{equation}\label{4mom}
 p = (0, 0, \omega, \omega)
\end{equation}
of a massless particle, the momentum remains invariant.  It therefore
satisfies the condition for the little group.  If we apply this matrix
to the electromagnetic four-potential
\begin{equation}
 A = (A_{1}, A_{2}, A_{3}, A_{0}) \exp{(i(kz -\omega t))} ,
\end{equation}
with $A_{3} = A_{0}$ which is the Lorentz condition,
the result is a gauge transformation.  This is what Janner and Janssen
discovered in their 1971 and 1972 papers~\cite{janner71}. Thus
the matrices $N_{1}$ and $N_{2}$ generate gauge transformations.

\section{Contraction of O(3) to E(2)}\label{o3e2}
In this section, we explain what the $E(2)$ group is.  We then
explain how we can obtain this group from the three-dimensional
rotation group by making a flat-surface or cylindrical approximation.
This contraction procedure will give a clue to obtaining the $E(2)$-like
symmetry for massless particles from the $O(3)$-like symmetry for
massive particles by taking the infinite-momentum limit.

The $E(2)$ transformations consist of a rotation and two translations on
a flat plane.  Let us start with the  rotation matrix applicable to
the column vector $(x, y, 1)$:
\begin{equation}\label{rot}
 R(\theta) = \pmatrix{\cos\theta & -\sin\theta & 0 \cr
\sin\theta & \cos\theta & 0 \cr 0 & 0 & 1} .
\end{equation}
Let us then consider the translation matrix:
\begin{equation}
 T(a, b) = \pmatrix{1 & 0 & a \cr 0 & 1 & b \cr 0 & 0 & 1} .
\end{equation}
If we take the product $T(a, b) R(\theta)$,
\begin{equation}\label{eucl}
E(a, b, \theta) = T(a, b) R(\theta)
= \pmatrix{\cos\theta & -\sin\theta & a \cr
\sin\theta & \cos\theta & b \cr 0 & 0 & 1} .
\end{equation}
This is the Euclidean transformation matrix applicable to the
two-dimensional $x y$ plane.  The matrices $R(\theta)$ and $T(a,b)$
represent the rotation and translation subgroups respectively.  The
above expression is not a direct product because $R(\theta)$ does not
commute with $T(a,b)$.  The translations constitute an Abelian invariant
subgroup because two different $T$ matrices commute with each other,
and because
\begin{equation}
 R(\theta) T(a,b) R^{-1}(\theta) = T(a',b') .
\end{equation}
The rotation subgroup is not invariant because the conjugation
\begin{equation}
 T(a,b) R(\theta) T^{-1}(a,b)
\end{equation}
does not lead to another rotation.

We can write the above transformation matrix in terms of generators.
The rotation is generated by
\begin{equation}
 J_{3} = \pmatrix{0 & -i & 0 \cr i & 0 & 0 \cr 0 & 0 & 0} .
\end{equation}
The translations are generated by
\begin{equation}
 P_{1} = \pmatrix{0 & 0 & i \cr 0 & 0 & 0 \cr 0 & 0 & 0} , \qquad
P_{2} = \pmatrix{0 & 0 & 0 \cr 0 & 0 & i \cr 0 & 0 & 0} .
\end{equation}
These generators satisfy the commutation relations:
\begin{equation}\label{e2com}
[P_{1}, P_{2}] = 0 , \qquad [J_{3}, P_{1}] = iP_{2} , \qquad
[J_{3}, P_{2}] = -iP_{1} .
\end{equation}
This $E(2)$ group is not only convenient for illustrating the groups
containing an Abelian invariant subgroup, but also occupies an
important place in constructing representations for the little
group for massless particles, since the little group for massless
particles is locally isomorphic to the above $E(2)$ group.

The contraction of $O(3)$ to $E(2)$ is well known and is often called
the Inonu-Wigner contraction~\cite{inonu53}.  The question is whether
the $E(2)$-like little group can be obtained from the $O(3)$-like
little group.  In order to answer this question, let us closely look
at the original form of the Inonu-Wigner contraction.  We start with
the generators of $O(3)$.  The $J_{3}$ matrix is given in Eq.(\ref{j3}),
and
\begin{equation}\label{o3gen}
J_{2} = \pmatrix{0&0&i\cr0&0&0\cr-i&0&0} ,  \qquad
J_{3} = \pmatrix{0&-i&0\cr i &0&0\cr0&0&0} .
\end{equation}
The Euclidean group $E(2)$ is generated by $J_{3}, P_{1}$ and $P_{2}$,
and their Lie algebra has been discussed in Sec.~\ref{intro}.

\begin{figure}[thb]  
\centerline{\psfig{figure=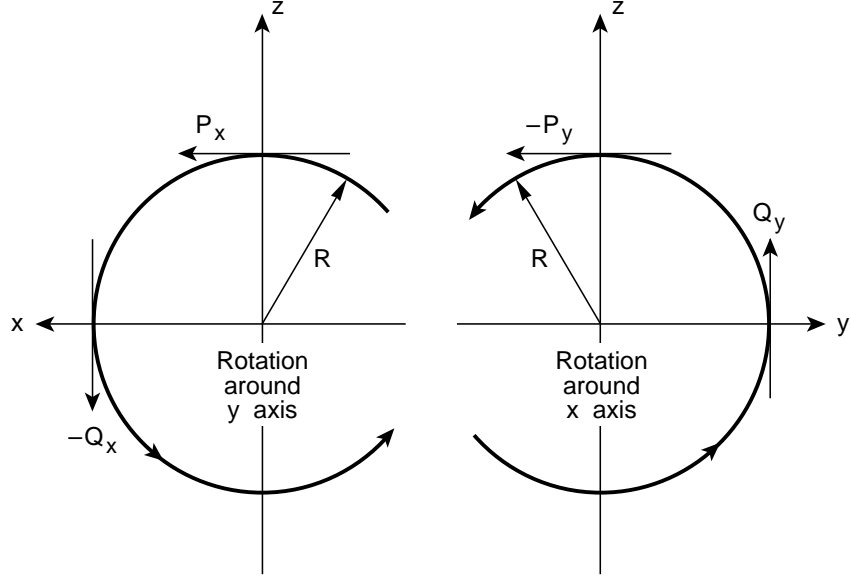,angle=0,height=80mm}}
\caption{North-pole and Equatorial-belt approximations.  The north-pole
approximation leads to the contraction of O(3) to E(2).  The
equatorial-belt approximation leads corresponds to the contraction the
cylindrical group.}\label{equator}
\end{figure}

Let us transpose the Lie algebra of the $E(2)$ group.  Then $P_{1}$ and
$P_{2}$ become $Q_{1}$ and $Q_{2}$ respectively, where
\begin{equation}
Q_{1} = \pmatrix{0&0&0\cr0&0&0\cr i &0&0} , \qquad
Q_{2} = \pmatrix{0&0&0\cr0&0&0\cr0&i&0} .
\end{equation}
Together with $J_{3}$, these generators satisfy the
same set of commutation relations as that for
$J_{3}, P_{1}$, and $P_{2}$ given in Eq.(\ref{e2com}):
\begin{equation}
[Q_{1}, Q_{2}] = 0 , \qquad [J_{3}, Q_{1}] = iQ_{2} , \qquad
[J_{3}, Q_{2}] = -iQ_{1} .
\end{equation}
These matrices generate transformations of a point on a circular
cylinder.  Rotations around the cylindrical axis are generated by
$J_{3}$.  The matrices $Q_{1}$ and $Q_{2}$ generate translations along
the direction of $z$ axis.  The group generated by these three matrices
is called the {\it cylindrical group}~\cite{kiwi87jm,kiwi90jm}.

We can achieve the contractions to the Euclidean and cylindrical groups
by taking the large-radius limits of
\begin{equation}\label{inonucont}
 P_{1} = {1\over R} B^{-1} J_{2} B ,
\qquad P_{2} = -{1\over R} B^{-1} J_{1} B ,
\end{equation}
and
\begin{equation}
 Q_{1} = -{1\over R}B J_{2}B^{-1} , \qquad
Q_{2} = {1\over R} B J_{1} B^{-1} ,
\end{equation}
where
\begin{equation}\label{bmatrix}
 B(R) = \pmatrix{1&0&0\cr0&1&0\cr0&0&R}  .
\end{equation}
The vector spaces to which the above generators are applicable are
$(x, y, z/R)$ and $(x, y, Rz)$ for the Euclidean and cylindrical groups
respectively.  They can be regarded as the north-pole and equatorial-belt
approximations of the spherical surface respectively~\cite{kiwi87jm}.
Fig.~\ref{equator} illustrates how the Euclidean and cylindrical
contractions are made.

\section{Contraction of O(3)-like Little Group to E(2)-like Little
Group}\label{contrac}

Since $P_{1} (P_{2})$ commutes with $Q_{2} (Q_{1})$, we can consider the
following combination of generators.
\begin{equation}
 F_{1} = P_{1} + Q_{1} , \qquad F_{2} = P_{2} + Q_{2} .
\end{equation}
Then these operators also satisfy the commutation relations:
\begin{equation}\label{commuf}
[F_{1}, F_{2}] = 0 , \qquad [J_{3}, F_{1}] = iF_{2} ,  \qquad
[J_{3}, F_{2}] = -iF_{1} .
\end{equation}
However, we cannot make this addition using the three-by-three matrices
for $P_{i}$ and $Q_{i}$ to construct three-by-three matrices for $F_{1}$
and $F_{2}$, because the vector spaces are different for the $P_{i}$ and
$Q_{i}$ representations.  We can accommodate this difference by creating
two different $z$ coordinates, one with a contracted $z$ and the other
with an expanded $z$, namely $(x, y, Rz, z/R)$.  Then the generators
become
\begin{equation}
P_{1} = \pmatrix{0&0&0&i\cr0&0&0&0\cr0&0&0&0\cr0&0&0&0} ,  \qquad
P_{2} = \pmatrix{0&0&0&0\cr0&0&0&i\cr0&0&0&0\cr0&0&0&0} ,
\end{equation}
and
\begin{equation}
Q_{1} = \pmatrix{0&0&0&0\cr0&0&0&0\cr i &0&0&0\cr0&0&0&0} ,  \qquad
Q_{2} = \pmatrix{0&0&0&0\cr0&0&0&0\cr0&i&0&0\cr0&0&0&0} .
\end{equation}
Then $F_{1}$ and $F_{2}$ will take the form
\begin{equation}\label{f1f2}
F_{1} = \pmatrix{0&0&0&i\cr0&0&0&0\cr i &0&0&0\cr0&0&0&0} ,  \qquad
F_{2} = \pmatrix{0&0&0&0\cr0&0&0&i\cr0&i&0&0\cr0&0&0&0} .
\end{equation}
The rotation generator $J_{3}$ takes the form of Eq.(\ref{j3}).
These four-by-four matrices satisfy the E(2)-like commutation relations
of Eq.(\ref{commuf}).

\begin{figure}[thb]  
\centerline{\psfig{figure=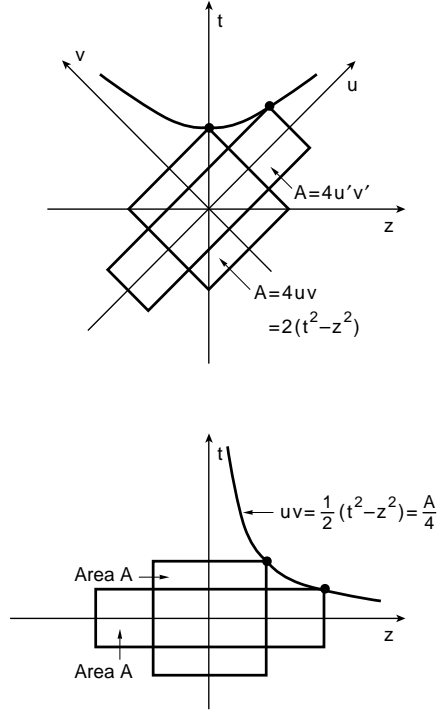,angle=0,height=95mm}}
\caption{Light-cone coordinates.  When the system is Lorentz-boosted,
one of the axes expands while the other becomes contracted.  Both
the expansion and contraction are needed for the contraction of
the $O(3)$-like little group to $E(2)$-like little group.}\label{lcone}
\end{figure}

Now the $B$ matrix of Eq.(\ref{bmatrix}), can be expanded to
\begin{equation}\label{bmatrix2}
 B(R) = \pmatrix{1&0&0&0\cr0&1&0&0\cr0&0&R&0\cr0&0&0&1/R} .
\end{equation}
This matrix includes both the contraction and expansion in the light-cone
coordinate system, as illustrated in Fig.~\ref{lcone}.
If we make a similarity transformation on the above form using the matrix
\begin{equation}\label{simil}
 \pmatrix{1&0&0&0\cr0&1&0&0\cr0&0&1/\sqrt{2} &-1/\sqrt{2}
\cr0&0&1/\sqrt{2}&1/\sqrt{2}} ,
\end{equation}
which performs a 45-degree rotation of the third and fourth coordinates,
then this matrix becomes
\begin{equation}\label{simil2}
 \pmatrix{1&0&0&0\cr0&1&0&0\cr0&0 & \cosh\eta & \sinh\eta
\cr0 & 0 & \sinh\eta & \cosh\eta} ,
\end{equation}
with $R = e^\eta$.  This form is the Lorentz boost matrix along the $z$
direction.  If we start with the set of expanded rotation generators
$J_{3}$ of Eq.(\ref{j3}), and
perform the same operation as the original Inonu-Wigner contraction
given in Eq.(\ref{inonucont}), the result is
\begin{equation}
 N_{1} = {1\over R} B^{-1} J_{2} B ,
\qquad N_{2} = -{1\over R} B^{-1} J_{1} B ,
\end{equation}
where $N_{1}$ and $N_{2}$ are given in Eq.(\ref{n1n2}).  The generators
$N_{1}$ and $N_{2}$ are the contracted $J_{2}$ and $J_{1}$ respectively
in the infinite-momentum and/or zero-mass limit.

It was noted in Sec.~\ref{gauge} that $N_{1}$ and $N_{2}$ generate
gauge transformations on massless particles.  Thus the contraction of
the transverse rotations leads to gauge transformations.

We have seen in this section that Wigner's $O(3)$-like little group
can be contracted into the $E(2)$-like little group for massless
particles.  Here, we worked out explicitly for the spin-1 case, but
this mechanism should be applicable to all other spins.  Of particular
interest is spin-1/2 particles.  This has been studied by Han, Kim
and Son~\cite{hks82}.  They noted that there are also gauge
transformations for spin-1/2 particles, and the polarization of
neutrinos is a consequence of gauge invariance.  It has also been
shown that the gauge dependence of spin-1 particles can be traced to
the gauge variable of the spin-1/2 particle~\cite{hks86jm}.  It would
be very interesting to see how the present formalism is applicable to
higher-spin particles.

Another case of interest is the space-time symmetry of relativistic
extended particles.  In 1973~\cite{kn73}, the present authors constructed
a ground-state harmonic oscillator wave function which can be
Lorentz-boosted.  It was later found that this oscillator formalism
can be extended to represent the $O(3)$-like little
group~\cite{kno79,knp86}.  This oscillator formalism has a stormy
history because it ultimately plays a pivotal role in combining
quantum mechanics and special relativity~\cite{dir45,yuka53}.

With these wave functions, we propose to solve the following problem in
high-energy physics.  The quark model works well when hadrons are at
rest or move slowly.  However, when they move with speed close to that
of light, they appear as a collection of an infinite-number of
partons~\cite{fey69}.  The question then is whether the parton model is
a Lorentz-boosted quark model.  This question has been addressed
before~\cite{kn77par,kim89}, but it can generate more interesting
problems~\cite{kiwi90pl}.  The present situation is presented in the
Table~\ref{table1}.

\begin{table}[thb]

\caption{Massive and massless particles in one package.  Wigner's
little group unifies the internal space-time symmetries for massive and
massless particles.  It is a great challenge for us to find
another unification: the unification of the quark and parton pictures in
high-energy physics.}\label{table1}

\vspace{3mm}

\begin{center}

\begin{tabular}{lccc}

\hline
{}&{}&{}&{}\\
{} & Massive, Slow \hspace{6mm} & COVARIANCE \hspace{6mm}&
Massless, Fast \\[4mm]\hline
{}&{}&{}&{}\\
Energy- & {}  & Einstein's & {} \\
Momentum & $E = p^{2}/2m$ & $ E = [p^{2} + m^{2}]^{1/2}$ & $E = p$
\\[4mm]\hline
{}&{}&{}&{}\\
Internal & $S_{3}$ & {}  &  $S_{3}$ \\[-1mm]
Space-time &{} & Wigner's  & {} \\ [-1mm]
Symmetry & $S_{1}, S_{2}$ & Little Group & Gauge Trans. \\[4mm]\hline
{}&{}&{}&{}\\
Relativistic & {} & One  &  {} \\[-1mm]
Extended & Quark Model & Covariant  & Parton Model\\ [-1mm]
Particles & {} & Theory &{} {} \\[4mm]\hline

\end{tabular}

\end{center}
\end{table}


We are now ready to consider the third row of Table~\ref{table1}.
In the this table, we would like to say that the quark model
and the parton model are two different manifestation of one covariant
entity.  In order to appreciate fully this covariant aspect, let us
examine Feynman's style of doing physics.

\section{Feynman's World}\label{feynm}
Feynman was quite fond of using harmonic oscillators to probe new
territories of physics.  In this section, we examine which
oscillator formalism is most suitable to interpret some of
Feynman's papers during the period 1969 -- 1972.  This formalism
should accommodate special relativity and quantum mechanics of
extended objects.

Let us start with a simple physical system.
Two coupled harmonic oscillators play many important roles in physics.
In group theory, it generates symmetry group as rich as
$O(3,3)$~\cite{hkn95jm}.  It has many interesting subgroups useful
in all branches of physics.  The group $O(3,1)$ is of course essential
for studying covariance in special relativity.  It is applicable to
three space-like variables and one time-like variable.  In the
harmonic oscillator regime, those
three space-like coordinates are separable.  Thus, it is possible to
separate longitudinal and transverse coordinates.  If we leave out
the transverse coordinates which do not participate in Lorentz boosts,
the only relevant variables are longitudinal and time-like variables.
The symmetry group for this case is $O(1,1)$ easily derivable from
the Hamiltonian for the two-oscillator system.

It is widely known that this simple mathematical device is the basic
language for two-photon coherent states known as squeezed states of
light~\cite{yuen76,knp91}.
However, this $O(1,1)$ device plays a much more powerful role in physics.
According to Feynman, {\it the adventure of our science of physics is a
perpetual attempt to recognize that the different aspects of nature
are really different aspects of the same thing}~\cite{feyaip}.  Feynman
wrote many papers on different subjects of physics, but they are
coming from one paper according to him.  We are not able to combine
all of his papers, but we can consider three of his papers published
during the period 1969-72.

In this paper, we would like to consider Feynman's 1969 report on
partons~\cite{fey69}, the 1971 paper he published with his students
on the quark model based on harmonic oscillators~\cite{fkr71}, and
the chapter on density matrix in his
1972 book on statistical mechanics~\cite{fey72}.  In these three
different papers, Feynman deals with three distinct aspects of nature.
We shall see whether Feynman was saying the same thing in these
papers. For this purpose, we shall use the $O(1,1)$ symmetry derivable
directly from the Hamiltonian for two coupled
oscillators~\cite{hkn99ajp}.  The standard procedure for this
two-oscillator system is to separate the Hamiltonian using normal
coordinates.  The transformation to the normal coordinate system
becomes very simple if the two oscillators are identical.  We shall
use this simple mathematics to find a common ground for the
above-mentioned articles written by Feynman.

First, let us look at Feynman's book on statistical
mechanics~\cite{fey72}.  He makes the following statement about the
density matrix. {\it When we solve a quantum-mechanical problem, what we
really do is divide the universe into two parts - the system in which we
are interested and the rest of the universe.  We then usually act as if
the system in which we are interested comprised the entire universe.
To motivate the use of density matrices, let us see what happens when we
include the part of the universe outside the system}.

In order to see clearly what Feynman had in mind, we use the
above-mentioned couples oscillators.  One of the oscillators is the
world in which we are interested and the other oscillator as the rest
of the universe.  There will be no effects on the first oscillator if
the system is decoupled.  Once coupled, we need a normal coordinate
system in order separate the Hamiltonian.  Then it is straightforward
to write down the wave function of the system.  Then the mathematics
of this oscillator system is directly applicable to Lorentz-boosted
harmonic oscillator wave functions, where one variable is the
longitudinal coordinate and the other is the time variable.  The system
is uncoupled if the oscillator wave function is at rest, but the
coupling becomes stronger as the oscillator is boosted to a high-speed
Lorentz frame~\cite{knp86}.

We shall then note that for two-body system, such as the hydrogen atom,
there is a time-separation variable which is to be linearly mixed with
the longitudinal space-separation variable.  This space-separation
variable is known as the Bohr radius, but we never talk about the
time-separation variable in the present form of quantum mechanics,
because this time-separation variable belongs to Feynman's rest of the
universe.

If we pretend not to know this time-separation variable, the entropy
of the system will increase when the oscillator is boosted to a
high-speed system~\cite{kiwi90pl}.  Does this increase in entropy
correspond to decoherence?  Not necessarily.  However, in 1969, Feynman
observed the parton effect in which a rapidly moving hadron appears
as a collection of incoherent partons~\cite{fey69}.  This is the
decoherence mechanism of current interest.

\section{Two Coupled Oscillators}\label{coupled}
Two coupled harmonic oscillators serve many different purposes in
physics.  It is well known that this oscillator problem can be
formulated into a problem of a quadratic equation in two variables.
To make a long story short, let us consider a system of two identical
oscillators coupled together by a spring.  The Hamiltonian is
\begin{equation}\label{hamil2}
H = {1\over 2m}\left\{p^{2}_{1} + p^{2}_{2} \right\} +
{1\over 2}\left\{K \left(x_{1}^{2} + x^{2}_{2} \right)
+ 2C x_{1} x_{2} \right\} .
\end{equation}
We are now ready to decouple this Hamiltonian by
making the coordinate rotation:
\begin{equation}\label{normal}
y_{1} = {1 \over \sqrt{2}} \left(x_{1}  - x_{2} \right) , \qquad
y_{2} = {1 \over \sqrt{2}} \left(x_{1}  + x_{2} \right) .
\end{equation}
In terms of this new set of variables, the Hamiltonian can be written as
\begin{equation}\label{eq.6}
H = {1\over 2m} \left\{p^{2}_{1} + p^{2}_{2} \right\} +
{K\over 2}\left\{e^{2\eta} y^{2}_{1} + e^{-2\eta} y^{2}_{2} \right\} ,
\end{equation}
with
\begin{equation}\label{omega}
\exp{(\eta)} = \sqrt{(K + C)/(K - C)} .
\end{equation}
Thus $\eta$ measures the strength of the coupling.
If $y_{1}$ and $y_{2}$ are measured in units of $(mK)^{1/4} $,
the ground-state wave function of this oscillator system is
\begin{equation}\label{wfc}
\psi_{\eta}(x_{1},x_{2}) = \left({1 \over {\pi}}\right)^{1/2}
\exp{\left\{-{1\over 2}(e^{\eta} y^{2}_{1} + e^{-\eta} y^{2}_{2})
\right\} } .
\end{equation}
The wave function is separable in the $y_{1}$ and $y_{2}$ variables.
However, for the variables $x_{1}$ and $x_{2}$, the story is quite
different.

The key question is how quantum mechanical calculations in the world
of the observed variable are affected when we average over the other
variable.  The $x_{2}$ space in this case corresponds to Feynman's
rest of the universe, if we only consider quantum mechanics in the
$x_{1}$ space.  As was discussed in the literature for several
different purposes~\cite{knp91,knp86}, the wave function of
Eq.(\ref{wfc}) can be expanded as
\begin{equation}\label{expan}
\psi_{\eta }(x_{1},x_{2}) = {1 \over \cosh\eta}\sum^{}_{k}
\left(\tanh{\eta \over 2}\right)^{k} \phi_{k}(x_{1}) \phi_{k}(x_{2}) .
\end{equation}
This expansion corresponds to the two-photon coherent states in Yuen's
paper~\cite{yuen76}, and the wave function of Eq.(\ref{wfc}) is
a squeezed wave function~\cite{knp91}.

The question then is what lessons we can learn from the situation in
which we average over the $x_{2}$ variable.
In order to study this problem, we use the density matrix.  From this
wave function, we can construct the pure-state density matrix
\begin{equation}
\rho(x_{1},x_{2};x_{1}',x_{2}')
= \psi_{\eta}(x_{1},x_{2})\psi_{\eta}(x_{1}',x_{2}') ,
\end{equation}
If we are not able to make observations on the $x_{2}$, we should
take the trace of the $\rho$ matrix with respect to the $x_{2}$
variable.  Then the resulting density matrix is
\begin{equation}\label{integ}
\rho(x, x') = \int \psi_{\eta}(x,x_{2})
\left\{\psi_{\eta}(x',x_{2})\right\}^{*} dx_{2} .
\end{equation}
We have simplicity replaced $x_{1}$ and $x'_{1}$ by $x$ and $x'$
respectively.
If we perform the integral of Eq.(\ref{integ}), the result is
\begin{equation}\label{dmat}
\rho(x,x') = \left({1 \over \cosh(\eta/2)}\right)^{2}
\sum^{}_{k} \left(\tanh{\eta \over 2}\right)^{2k}
\phi_{k}(x)\phi^{*}_{k}(x') ,
\end{equation}
which leads to $Tr(\rho) = 1$.  It is also straightforward to compute
the integral for $Tr(\rho^{2})$.  The calculation leads to
\begin{equation}
Tr\left(\rho^{2} \right)
= \left({1 \over \cosh(\eta/2)}\right)^{4}
\sum^{}_{k} \left(\tanh{\eta \over 2}\right)^{4k} .
\end{equation}
The sum of this series is $(1/\cosh\eta)$, which is smaller than one
if the parameter $\eta$ does not vanish.

This is of course due to the fact that we are averaging over the $x_{2}$
variable which we do not measure.  The standard way to measure this
ignorance is to calculate the entropy defined as
\begin{equation}
S = - Tr\left(\rho \ln(\rho) \right) ,
\end{equation}
where $S$ is measured in units of Boltzmann's constant.  If we use the
density matrix given in Eq.(\ref{dmat}), the entropy becomes
\begin{equation}
S = 2 \left\{\cosh^{2}\left({\eta \over 2}\right)
\ln\left(\cosh{\eta \over 2}\right) -
\sinh^{2}\left({\eta \over 2}\right)
\ln\left(\sinh{\eta \over 2} \right)\right\} .
\end{equation}
This expression can be translated into a more familiar form if
we use the notation
\begin{equation}
\tanh{\eta \over 2} = \exp\left(-{\hbar\omega \over kT}\right) ,
\end{equation}
where $\omega$ is given in Eq.(\ref{omega})~\cite{hkn89pl}.

It is known in the literature that this rise in entropy and temperature
causes the Wigner function to spread wide in phase space causing an
increase of uncertainty~\cite{hkn99ajp}.  Certainly, we cannot reach a
classical limit by increasing the uncertainty.  On the other hand, we
are accustomed to think this entropy increase has something to do with
decoherence, and we are also accustomed to think the lack of coherence
has something to do with a classical limit.  Are they compatible?  We
thus need a new vision in order to define precisely the word
``decoherence.''

\section{Time-separation Variable in Feynman's Rest of
the Universe}\label{restof}
Quantum field theory has been quite successful in terms of
perturbation techniques in quantum electrodynamics.  However, this
formalism is basically based
on the S matrix for scattering problems and useful only for physically
processes where a set of free particles becomes another set of free
particles after interaction.  Quantum field theory does not address
the question of localized probability distributions and their
covariance under Lorentz transformations.
The Schr\"odinger quantum mechanics of the hydrogen atom deals with
localized probability distribution.  Indeed, the localization condition
leads to the discrete energy spectrum.  Here, the uncertainty relation
is stated in terms of the spatial separation between the proton and
the electron.  If we believe in Lorentz covariance, there must also
be a time separation between the two constituent particles.

Before 1964~\cite{gell64}, the hydrogen atom was used for
illustrating bound states.  These days, we use hadrons which are
bound states of quarks.  Let us use the simplest hadron consisting of
two quarks bound together with an attractive force, and consider their
space-time positions $x_{a}$ and $x_{b}$, and use the variables
\begin{equation}
X = (x_{a} + x_{b})/2 , \qquad x = (x_{a} - x_{b})/2\sqrt{2} .
\end{equation}
The four-vector $X$ specifies where the hadron is located in space and
time, while the variable $x$ measures the space-time separation
between the quarks.  According to Einstein, this space-time separation
contains a time-like component which actively participates as can be
seen from
\begin{equation}\label{boostm}
\pmatrix{z' \cr t'} = \pmatrix{\cosh \eta & \sinh \eta \cr
\sinh \eta & \cosh \eta } \pmatrix{z \cr t} ,
\end{equation}
when the hadron is boosted along the $z$ direction.
In terms of the light-cone variables defined as~\cite{dir49}
\begin{equation}
u = (z + t)/\sqrt{2} , \qquad v = (z - t)/\sqrt{2} .
\end{equation}
The boost transformation of Eq.(\ref{boostm}) takes the form
\begin{equation}\label{lorensq}
u' = e^{\eta } u , \qquad v' = e^{-\eta } v .
\end{equation}
The $u$ variable becomes expanded while the $v$ variable becomes
contracted.

Does this time-separation variable exist when the hadron is at rest?
Yes, according to Einstein.  In the present form of quantum mechanics,
we pretend not to know anything about this variable.  Indeed, this
variable belongs to Feynman's rest of the universe.  In this report,
we shall see the role of this time-separation variable in decoherence
mechanism.

\begin{figure}
\centerline{\psfig{figure=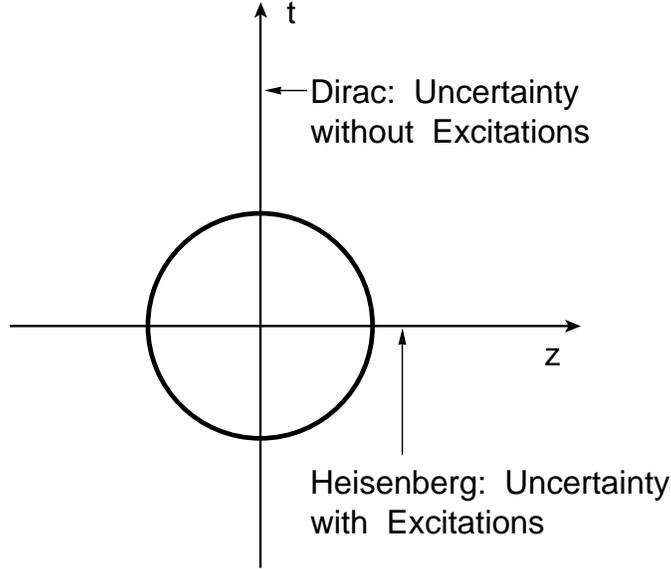,angle=0,height=80mm}}
\vspace{5mm}
\caption{Space-time picture of quantum mechanics.  There
are quantum excitations along the space-like longitudinal direction, but
there are no excitations along the time-like direction.  The time-energy
relation is a c-number uncertainty relation.}\label{quantum}
\end{figure}

Also in the present form of quantum mechanics, there is an uncertainty
relation between the time and energy variables.  However, there are
no known time-like excitations.  Unlike Heisenberg's
uncertainty relation applicable to position and momentum, the time and
energy separation variables are c-numbers, and we are not allowed to
write down the commutation relation between them.  Indeed, the
time-energy uncertainty relation is a c-number uncertainty
relation~\cite{dir27}, as is illustrated in Fig.~\ref{quantum}

How does this space-time asymmetry fit into the world of
covariance~\cite{kn73}.  This question was
studied in depth by the present authors.  The
answer is that Wigner's $O(3)$-like little group is not a
Lorentz-invariant symmetry, but is a covariant symmetry~\cite{wig39}.
It has been shown that the time-energy uncertainty applicable to the
time-separation variable fits perfectly into the $O(3)$-like symmetry
of massive relativistic particles~\cite{knp86}.

The c-number time-energy uncertainty relation allows us to write down
a time distribution function without excitations~\cite{knp86}.
If we use Gaussian forms for both space and time distributions, we
can start with the expression
\begin{equation}\label{ground}
\left({1 \over \pi} \right)^{1/2}
\exp{\left\{-{1 \over 2}\left(z^{2} + t^{2}\right)\right\}}
\end{equation}
for the ground-state wave function.  What do Feynman {\it et al.}
say about this oscillator wave function?

\begin{figure}
\centerline{\psfig{figure=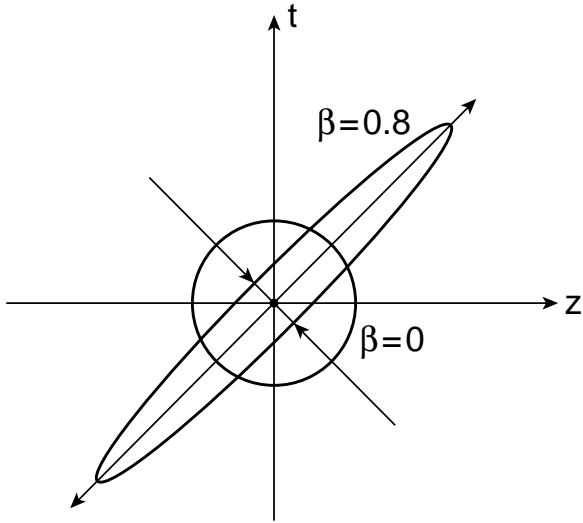,angle=0,height=80mm}}
\caption{Effect of the Lorentz boost on the space-time
wave function.  The circular space-time distribution at the rest frame
becomes Lorentz-squeezed to become an elliptic
distribution.}\label{ellipse}
\end{figure}

In his classic 1971 paper~\cite{fkr71}, Feynman {\it et al.} start
with the following Lorentz-invariant differential equation.
\begin{equation}\label{osceq}
{1\over 2} \left\{x^{2}_{\mu} -
{\partial^{2} \over \partial x_{\mu }^{2}}
\right\} \psi(x) = \lambda \psi(x) .
\end{equation}
This partial differential equation has many different solutions
depending on the choice of separable variables and boundary conditions.
Feynman {\it et al.} insist on Lorentz-invariant solutions which are
not normalizable.  On the other hand, if we insist on normalization,
the ground-state wave function takes the form of Eq.(\ref{ground}).
It is then possible to construct a representation of the
Poincar\'e group from the solutions of the above differential
equation~\cite{knp86}.  If the system is boosted, the wave function
becomes
\begin{equation}\label{eta}
\psi_{\eta }(z,t) = \left({1 \over \pi }\right)^{1/2}
\exp\left\{-{1\over 2}\left(e^{-2\eta }u^{2} +
e^{2\eta}v^{2}\right)\right\} .
\end{equation}
This wave function becomes Eq.(\ref{ground}) if $\eta$ becomes zero.
The transition from Eq.(\ref{ground}) to Eq.(\ref{eta}) is a
squeeze transformation.  The wave function of Eq.(\ref{ground}) is
distributed within a circular region in the $u v$ plane, and thus
in the $z t$ plane.  On the other hand, the wave function of
Eq.(\ref{eta}) is distributed in an elliptic region with the light-cone
axes as the major and minor axes respectively.  If $\eta$ becomes very
large, the wave function becomes concentrated along one of the
light-cone axes.  Indeed, the form given in Eq.(\ref{eta}) is a
Lorentz-squeezed wave  function.  This squeeze mechanism is
illustrated in Fig.~\ref{ellipse}.

It is interesting to note that the Lorentz-invariant differential
equation of Eq.(\ref{osceq}) contains the time-separation variable
which belongs to Feynman's rest of the universe.  Furthermore, the
wave function of Eq.(\ref{ground}) is identical to that of
Eq.(\ref{wfc}) for the coupled oscillator system, if the variables
$z$ and $t$ are replaced $x_{1}$ and $x_{2}$ respectively.
Thus the entropy increase due to the unobservable $x_{2}$ variable is
applicable to the unobserved time-separation variable $t$.


\section{Feynman's Parton Picture}\label{par}

It is a widely accepted view that hadrons are quantum bound states
of quarks having localized probability distribution.  As in all
bound-state cases, this localization condition is responsible for
the existence of discrete mass spectra.  The most convincing evidence
for this bound-state picture is the hadronic mass spectra which are
observed in high-energy laboratories~\cite{fkr71,knp86}.
However, this picture of bound states is applicable only to observers
in the Lorentz frame in which the hadron is at rest.  How would the
hadrons appear to observers in other Lorentz frames?  To answer this
question, can we use the picture of Lorentz-squeezed hadrons discussed
in Sec.~\ref{restof}.

The radius of the proton is $10^{-5}$ of that of the hydrogen atom.
Therefore, it is not unnatural to assume that the proton has a point
charge in atomic physics.  However, while carrying out experiments on
electron scattering from proton targets, Hofstadter in 1955 observed
that the proton charge is spread out~\cite{hofsta55}.
In this experiment, an electron emits a virtual photon, which
then interacts with the proton.  If the proton consists of quarks
distributed within a finite space-time region, the virtual photon will
interact with quarks which carry fractional charges.  The scattering
amplitude will depend on the way in which quarks are distributed
within the proton.  The portion of the scattering amplitude which
describes the interaction between the virtual photon and the proton
is called the form factor.

Although there have been many attempts to explain this phenomenon
within the framework of quantum field theory, it is quite natural
to expect that the wave function in the quark model will describe
the charge distribution.  In high-energy experiments, we are dealing
with the situation in which the momentum transfer in the scattering
process is large.  Indeed, the Lorentz-squeezed wave functions lead
to the correct behavior of the hadronic form factor for large
values of the momentum transfer~\cite{fuji70}.

Furthermore, in 1969, Feynman observed that a fast-moving hadron
can be regarded as a collection of many ``partons'' whose properties
do not appear to be quite different from those of the
quarks~\cite{fey69}.  For example, the number of quarks inside a
static proton is three, while the number of partons in a rapidly
moving proton appears to be infinite.  The question then is how
the proton looking like a bound state of quarks to one observer
can appear different to an observer in a different Lorentz frame?
Feynman made the following systematic observations.

\begin{itemize}

\item[a.]  The picture is valid only for hadrons moving with
  velocity close to that of light.

\item[b.]  The interaction time between the quarks becomes dilated,
   and partons behave as free independent particles.

\item[c.]  The momentum distribution of partons becomes widespread as
   the hadron moves fast.

\item[d.]  The number of partons seems to be infinite or much larger
    than that of quarks.

\end{itemize}

\noindent Because the hadron is believed to be a bound state of two
or three quarks, each of the above phenomena appears as a paradox,
particularly b) and c) together.

In order to resolve this paradox, let us write down the
momentum-energy wave function corresponding to Eq.(\ref{eta}).
If the quarks have the four-momenta $p_{a}$ and $p_{b}$, we can
construct two independent four-momentum variables~\cite{fkr71}
\begin{equation}
P = p_{a} + p_{b} , \qquad q = \sqrt{2}(p_{a} - p_{b}) ,
\end{equation}
where $P$ is the total four-momentum and is thus the hadronic
four-momentum.  $q$ measures the four-momentum separation between
the quarks.  Their light-cone variables are
\begin{equation}\label{conju}
q_{u} = (q_{0} - q_{z})/\sqrt{2} ,  \qquad
q_{v} = (q_{0} + q_{z})/\sqrt{2} .
\end{equation}
The resulting momentum-energy wave function is
\begin{equation}\label{phi}
\phi_{\eta }(q_{z},q_{0}) = \left({1 \over \pi }\right)^{1/2}
\exp\left\{-{1\over 2}\left(e^{-2\eta}q_{u}^{2} +
e^{2\eta}q_{v}^{2}\right)\right\} .
\end{equation}
Because we are using here the harmonic oscillator, the mathematical
form of the above momentum-energy wave function is identical to that
of the space-time wave function.  The Lorentz squeeze properties of
these wave functions are also the same.  This aspect of the squeeze
has been exhaustively discussed in the
literature~\cite{knp86,kn77par,kim89}.

When the hadron is at rest with $\eta = 0$, both wave functions
behave like those for the static bound state of quarks.  As $\eta$
increases, the wave functions become continuously squeezed until
they become concentrated along their respective positive
light-cone axes.  Let us look at the z-axis projection of the
space-time wave function.  Indeed, the width of the quark distribution
increases as the hadronic speed approaches that of the speed of
light.  The position of each quark appears widespread to the observer
in the laboratory frame, and the quarks appear like free particles.

\begin{figure}
\centerline{\psfig{figure=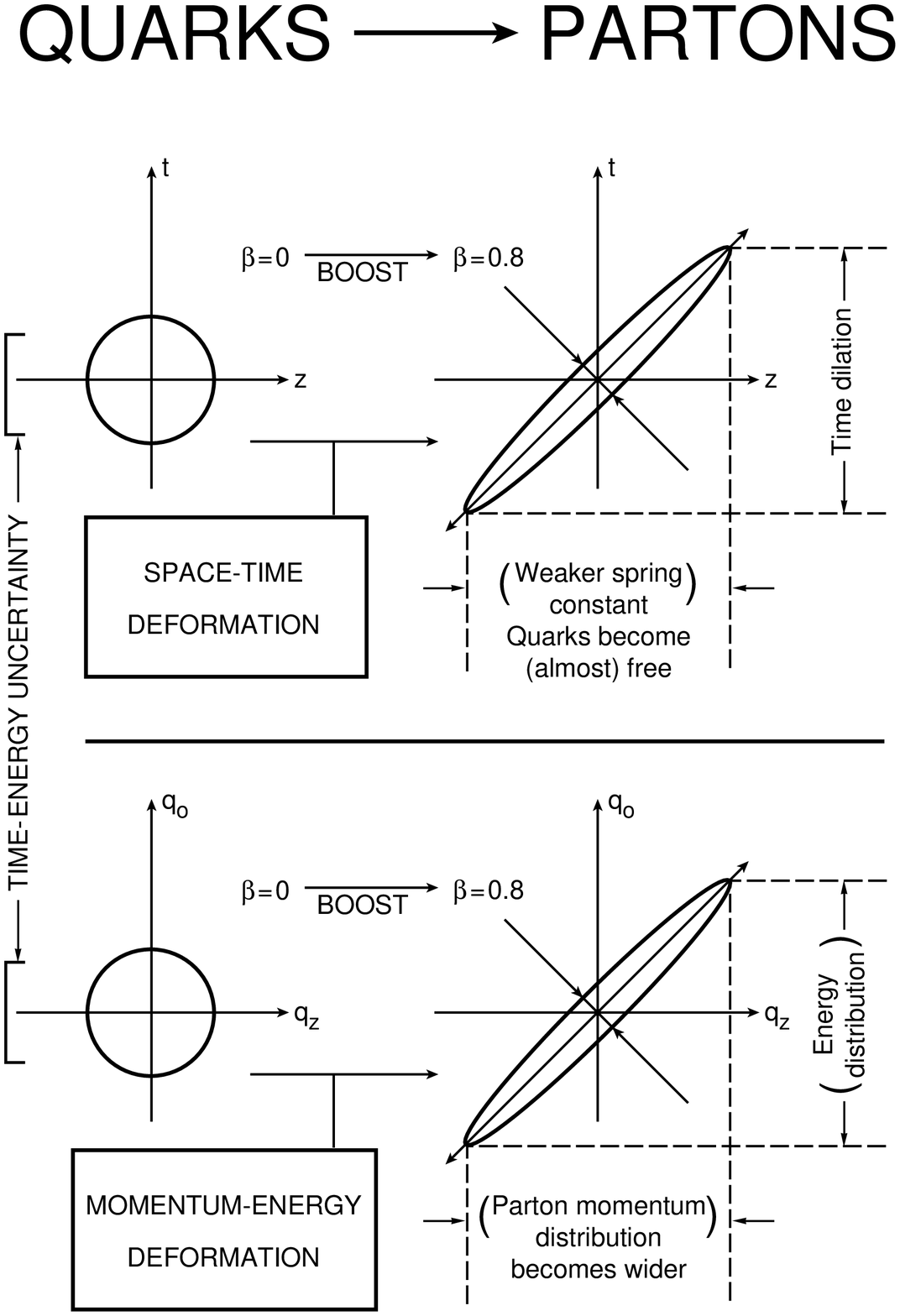,angle=0,height=100mm}}
\vspace{5mm}
\caption{Lorentz-squeezed space-time and momentum-energy
wave functions.  As the hadron's speed approaches that of light, both
wave functions become concentrated along their respective positive
light-cone axes.  These light-cone concentrations lead to Feynman's
 parton picture.}\label{parton}
\end{figure}

The momentum-energy wave function is just like the space-time wave
function, as is shown in Fig.~\ref{parton}.  The longitudinal momentum
distribution becomes wide-spread as the hadronic speed approaches the
velocity of light.  This is in contradiction with our expectation from
nonrelativistic quantum mechanics that the width of the momentum
distribution is inversely proportional to that of the position wave
function.  Our expectation is that if the quarks are free, they must
have their sharply defined momenta, not a wide-spread distribution.

However, according to our Lorentz-squeezed space-time and
momentum-energy wave functions, the space-time width and the
momentum-energy width increase in the same direction as the hadron
is boosted.  This is of course an effect of Lorentz covariance.
This indeed is the key to the resolution of the quark-parton
paradox~\cite{knp86,kn77par}.

The most puzzling problem in the parton picture is that partons in
the hadron appear as incoherent particles, while quarks are coherent
when the hadron is at rest.  Does this mean that the coherence is
destroyed by the Lorentz boost?   The answer is NO, and here is the
resolution to this puzzle.

When the hadron is boosted, the hadronic matter becomes squeezed and
becomes concentrated in the elliptic region along the positive
light-cone axis, as is illustrated in Figs.~\ref{ellipse} and
\ref{parton}.  The length of the major axis becomes expanded by
$e^{\eta}$, and the minor axis is contracted by $e^{\eta}$.

This means that the interaction time of the quarks among themselves
become dilated.  Because the wave function becomes wide-spread, the
distance between one end of the harmonic oscillator well and the
other end increases.  This effect, first noted by Feynman~\cite{fey69},
is universally observed in high-energy hadronic experiments.  The
period is oscillation increases like $e^{\eta}$.

On the other hand, the interaction time with the external signal,
since it is moving in the direction opposite to the direction of
the hadron, travels along the negative light-cone axis.
If the hadron contracts along the negative light-cone axis, the
interaction time decreases by $e^{-\eta}$.  The ratio of the interaction
time to the oscillator period becomes $e^{-2\eta}$.  The energy of each
proton coming out of the Fermilab accelerator is $900 GeV$.  This leads
the ratio to $10^{-6}$.  This is indeed a small number.  The external
signal is not able to sense the interaction of the quarks among
themselves inside the hadron.

Indeed, Feynman's parton picture is one concrete physical example
where the decoherence effect is observed.  As for the entropy, the
time-separation variable belongs to the rest of the universe.  Because
we are not able to observe this variable, the entropy increases
as the hadron is boosted to exhibit the parton effect.  The
decoherence is thus accompanied by an entropy increase.

Let us go back to the coupled-oscillator system.  The light-cone
variables in Eq.(\ref{eta}) correspond to the normal coordinates in
the coupled-oscillator system given in Eq.(\ref{normal}).  According
to Feynman's parton picture, the decoherence mechanism is determined
by the ratio of widths of the wave function along the two normal
coordinates.  The result is listed in the third row of
Table~\ref{table1}.

\section{Further Contents of Einstein's Formula for Energy, Mass,
and Momentum}\label{further}
In Table~\ref{table1}, we put Wigner's formalism and Feynman's
observation into a single package which could be called ``Further
Contents of Einstein's $E = mc^{2}$.''  To physicists, $E = mc^{2}$
means  $E = \sqrt{m^{2} + p^{2}}$.  Of course, the mass $m$ has
different meanings for these two different formulas, one for the
rest mass and the other for moving mass.  This distinction is so
obvious to physicists that there is a tendency not to mention it
in the physics literature.

However, the distinction is not so trivial to those who study how
special relativity was developed.  Indeed, there has been a recent
debate on this issue, and the debate is likely to continue.  However,
the present authors have not done enough research on this issue, but
would like to acknowledge a very comprehensive review article by
Okun~\cite{okun89}, entitled ``Concept of Mass'' and comments on
this article by various authors.

It is not clear whether Einstein was concerned with the question of
whether the particles are point particles or objects with internal
space-time structures, because the internal space-time symmetry was
not formulated until 1939 when Wigner published his paper based on
the little groups~\cite{wig39}.  On the otherhand, Wigner's approach
starts from Einstein's energy-momentum relation for free particles.
Thus, the energy-momentum relation remains valid for particles with
internal structure only if there is a covariant description of
internal space-time symmetries.  Indeed, this is the main point of
the present paper.

\begin{table}[thb]

\caption{Historical Necessity.  Newtonian Mechanics is
Galilean-covariant, and Maxwell's theory is Lorentz-covariant.
Do they have to be based on two different kind of covariance?
}\label{table2}

\vspace{3mm}

\begin{center}

\begin{tabular}{cccc}

\hline
{}&{}&{}&{}\\
{} & Galilean Covariance &{\hspace{20mm}}& Lorentz Covariance \\[4mm]\hline
{}&{}&{}&{}\\
Newtonian Mechanics & YES    &{}&  NO  \\[4mm]\hline
{}&{}&{}&{}\\
Maxwell Theory      &  NO    &{}&  YES   \\[4mm]\hline

\end{tabular}

\end{center}
\end{table}


Another historical question is who formulated the mathematics for
special relativity.  Here, prominent names are Lorentz, Poincar\'e and
Minkowski.  This is also an interesting and important issue.
The present authors have done some research along this line, but
not enough to make an impact on the existing literature.

The following point is well known, but seldom mentioned.  Before
Einstein, Newtonian mechanics and Maxwell's equations were based
on two different covariance principles, as is summarized in
Table~\ref{table2}.  Thus the development of special relativity
is of historical necessity to those, like Einstein, who believed
this world is one covariant world.

It is now firmly established that mechanics should also be
Lorentz-covariant.  Furthermore, it is a well-accepted view that
Lorentz covariance should become Galilean covariance for slow
particles.  This slow-speed limit is not as trivial as taking a
numerical limit of speed of particle divided by the speed of light.
This limiting process was worked out by Inonu and Wigner in their
1953 paper~\cite{inonu53}, where they introduced group contractions.
The Inonu-Wigner group contraction also allows us to take a
large-speed limit, which we used in the present paper.  It is
interesting to note that both limiting processes can be derived
from the Inonu-Wigner contraction.

\section*{Acknowledgments}
The author would like to thank Professor Lev Okun for sending us
his list of papers on the concept of mass, and also for helpful
comments.

\end{document}